# Analysis of energy, CO$_2$ emissions and economy of the technological migration for clean cooking in Ecuador

J. Martínez[a,b,d], Jaime Martí-Herrero[a,b,c,*], S. Villacís[a], A.J. Riofrio[a], D. Vaca[a]

[a] Instituto Nacional de Eficiencia Energética y Energías Renovables – INER. Dirección: Av. 6 de diciembre N33-32 e Ignacio Bossano, Edificio Torre Bossano, 2do. Piso Quito, Ecuador
[b] Universidad Internacional SEK Ecuador, Quito EC170134, Quito, Ecuador
[c] Centre Internacional de Mètodes Numèrics en Enginyeria (CIMNE), Building Energy and Environment Group, Edifici GAIA (TR14), C/Rambla Sant Nebridi 22, 08222 Terrassa, Barcelona, Spain
[d] Universidad de Alcalá, Departamento de teoría de la señal y comunicación, (Área de Ingeniería Mecánica) Escuela Politécnica, 28805 Alcalá de Henares, Madrid, España

ABSTRACT

The objective of this study is to analyze the CO$_2$ emissions and economic impacts of the implementation of the National Efficient Cooking Program (NECP) in Ecuador, which aims to migrate the population from Liquefied Petroleum Gas (LPG)-based stoves to electric induction stoves. This program is rooted in the current effort to change Ecuador's energy balance, with hydroelectric power expected to generate 83.61% of national electricity by 2022, ending the need for subsidized LPG. For this analysis, the 2014 baseline situation has been compared with two future scenarios for 2022: a business-as-usual scenario and an NECP-success scenario. This study demonstrates the viability of migration from imported fossil fuels to locally-produced renewable energy as the basis for an efficient cooking facility. The new policies scenario would save US$ 1.162 billion in annual government expenditure on cooking subsidies, and reducing CO$_2$ emissions associated to energy for cooking in 1.8 tCO$_2$/y.

## 1. Introduction

The General Assembly of the United Nations (UN) has declared 2014–2024 the "Decade of Sustainable Energy for All". Among other objectives, this resolution targets improved energy access for the approximately 1.3 billion people worldwide who still live without electricity, and the more than 2.6 billion people who rely on traditional biomass for cooking and heating (Smith, 2014).

As befits a problem of this scope, many programs have been developed not only to distribute and promote improved cooking facilities, but also to analyze the efficacy and design of the various strategies implemented. In China, for example, several programs to analyze improved household stoves have been carried out since the 1980s (Sinton et al., 2004). The objectives in Chinese programs were to delineate and evaluate the methods used to promote improved stoves, to assess the development of commercial stove production and marketing organizations (Sinton et al., 2004). Studies to understand household energy preferences for cooking in urban Ouagadougou have been undertaken to improve the clean cooking program in Burkina Faso (Ouedraogo et al., 2006). In the case of India, where energy for cooking uses accounts for a little over 80% of the total household energy consumption in rural areas (Purohit et al., 2002), the use of biogas plants, solar stoves and improved household biomass stoves for domestic cooking has been promoted (Purohit, et al., 2011). In Ethiopia, a technical document prepared with input from a sector-wide network mapped institutional factors that hamper the adoption of solar cookers in Africa (Kebede et al., 2014).

Clean cooking programs generally begin with the introduction of improved biomass stoves in rural households, advance to the promotion of improved charcoal stoves, and eventually attempt to transition to clean cooking fuels by promoting LPG stoves (Sinton et al., 2004; Shrimali, 2011). These programs have aimed to end the use of solid-fuel stoves with low energy efficiency and excessive emissions of toxic smoke, which primarily affects women and young children who spend significant time in the kitchen. Another historic motivator of these programs is that about 25% of outdoor particle pollution emissions, and significant contributions to CO$_2$ and shorter-lived greenhouse pollutants, are produced by the incomplete combustion and poor energy efficiency, characteristic of traditional biomass cooking combustion (Shen et al., 2011; Maes et al., 2012). With 3.9 million premature deaths annually attributed to traditional uses of biomass cooking fuels, these are now understood to be the largest single environmental health threat in the world, although they affect only about 40% of the world´s population (Smith et al., 2014).

As an alternative to the improve biomass-charcoal-LPG progression, several studies have been performed on clean cooking facilities using electricity. While promotion of electric coil resistance stoves has been cautioned against due to their low efficiencies and high power consumption (Smith and Sagar, 2014), other studies have observed that in areas where reliable power supply is available, electrical appliances such as electric kettles, rice cookers, ovens and microwaves are used and lead to a reduction in biomass used for cooking in an effective way (Smith and Sagar, 2014; Villacís et al., 2015).

In addition to electric appliance use, promotion of induction cooking is beginning in other areas. Recently, India's Himachal Pradesh state has developed a program for "Access to clean cooking alternatives in rural India". In nearly 4000 rural households induction stoves were introduced to improve clean cooking facilities (Banerjee et al., 2016). Although not a country-wide program, the availability of inexpensive portable induction cooking stoves is shifting India's population towards cooking with electricity. Although this transition is beginning mainly in urban areas due to greater electricity availability and lower power costs, these constraints are changing as electrification expands and prices for induction stoves fall (Smith et al., 2014). Energy management and greenhouse emissions reduction have been analyzed for the regional "Plan Fronteras" that introduced 5400 induction stoves in the north of Ecuador (Martínez-G et al., 2017). In that paper, is analyzed the effect on the electricity grid of the introduction of thousands of induction stoves. From regional experimental data, the research shows the need to increase the power in 1.3 times to cover the electricity demand peak due induction stoves introduction.

Ecuador is currently carrying out the National Efficient Cooking Program (NECP), which aims to migrate the nation's population from LPG stoves to induction stoves (CONELEC, 2015). This program is linked to a change in the national energy matrix, designed to take advantage of increasing national hydroelectricity production and reduce dependence on imported LPG. As the first program of its kind to promote national-level cooking fuel migration to electricity (with subsequent reductions in fossil fuel consumption and government subsidies for LPG), this program provides an interesting case study for analysis.

This paper analyzes the energy subsidy and user costs for cooking energy in the context of the NECP program in Ecuador. It analyzes the current situation and compares it with two future scenarios, one related to a business as usual scenario without NECP and the other considering it.

### 1.1. National Efficient Cooking Program (NECP)

The government of Ecuador currently subsidizes LPG for the general population for domestic use: a 15 kg bottle of LPG costs US $1.60 (official price), while in neighbor countries (Peru and Colombia) this price is on average thirteen times higher (US$17 and US$23, respectively) (Riofrio, 2015). The total cost of this subsidy to the government is about US$690 million per year, with approximately 5% of subsidized LPG lost to smuggling and 15% used for non-domestic propose (Martínez-Gómez, 2016). In addition, approximately 78% of Ecuador's bottled LPG is imported, which creates major dependency and a significant outflow of national funds abroad which considerably affects the balance of trade in Ecuador (CONELEC, 2013).

The NECP aims to transition around three million households actually using Liquefied Petroleum Gas (LPG) stoves to induction cooking, forming the first national-level program promoting this technology (MEER, 2013), (Villacís et al., 2015). With this program, Ecuador looks to replace the use of largely imported fossil fuels in its energy mix with locally produced renewable energies, investing US $11.62 billion in new hydroelectric power stations, and transmission infrastructure by 2022 (CONELEC, 2013). The Ecuadorian government estimates an investment of US$6.012 billion in hydropower plants, US $1.158 billion in improved transmission infrastructure, US$3.378 billion in household distribution, and US$1.071 billion in the NECP and the National Efficient Electric Heater Program (CONELEC, 2015).

Current estimates for the NECP project 4.3 million LPG stoves will be replaced by induction stoves by 2022, with calculations from the Instituo Nacional de Estadisticas y Censos (INEC, 2010) projecting 1.2% annual population growth (with Ecuador's total population expected to increase from 15.3 million in 2011 to 19.8 million in 2030) and projecting average family size for the same period to reduce from 3.7 member per family in 2014 to 3.37 for 2022 (INEC, 2010), (CONELEC, 2015).

## 2. Experimental methodology

### 2.1. Primary energy from electricity production

To conduct the economic analysis, the quantity of primary energy produced, disaggregated by energy source, was initially determined. Electricity generation data for 2014 was taken as a base line, acquired from database publications of the Ecuadorian electricity sector's regulatory agency, the, "Agencia de Control y Regulación de la Electricidad" (ARCONEL, 2014). Using this data, a detailed analysis of every renewable and non-renewable energy source was calculated, with the purpose of determining the amount of energy generated and its contribution in relationship to overall production. Applying the Eq. (1) the primary energy consumption for electricity generation for 2014 was determined.

$$Pe = \sum E_i * C_i \qquad (1)$$

Where $Pe$ is the total primary energy related to electricity, $E_i$ is the electric energy produced by a specific source $i$, including renewable and not renewable sources, and $C_i$ is the conversion factor to primary energy of the source $i$ at the point of consumption. This factor is unique for each energy source. In order to determine the conversion factors ($C_i$), the suggested values from the Instituto para la Diversificación y Ahorro de la Energía of Spain (IDAE, 2012) were used. Table 1 shows the resultant values of primary energy for the electricity generation balance in Ecuador for 2014. In similar way, $CO_2$ emissions associated to each energy source has been calculated, using conversion factors from IDAE.

In addition, projections for Ecuador's 2022 energy balance have been calculated, considering the government's ambition to make hydropower plants the main source of electric generation by 2022. Data for these projections comes from the Electrification Master Plan 2013–2022 (CONELEC, 2013), considering a scenario where thermoelectric generation is nearly halved vs. its 2014 levels, hydroelectric generation is tripled and non-conventional renewable generation increases 11%. In Table 1 the disaggregated electricity generation and conversion to primary energy sources for 2014 and 2022 is presented. The same conversion factors for primary energy have been used in both 2014 and 2022 calculations. It is expected that total electricity production will increase by 75.7% over the 8-year period. This increase in electricity production is accompanied with a reduction in 4.7 million of $tCO_2$ emission in 2022 respect 2014.

### 2.2. Cooking energy consumption at household level

For LPG stoves, previous studies on Ecuadorian households show average final daily energy consumption of 8.85 kWh (31.86 MJ), giving an average monthly consumption of 265.39 kWh (955.4 MJ) (Riofrio, 2015; CONELEC, 2013), which translates into 17.6 kg of LPG per month at commercial compression levels. In Ecuador, LPG is commercialized in 15 kg cylinders, therefore the average household would be using 1.13 cylinders per month for its cooking activities. The reduction of family size by 9% is not considered significant for energy consumption for cooking, and therefore, for the purposes of this study, consumption by a typical family for 2014 and 2022 is projected to remain the same. With these considerations it is possible to estimate the primary energy consumption for cooking activities for an Ecuadorian family under current conditions and in 2022 with various cooking fuels. For LPG stoves a primary energy factor of 1.05 has been considered (IDAE, 2012), resulting in consumption of 278.66 kWh a month, both for 2014 and 2022. For the calculation of the induction stoves' factor, the primary energy depends on the electric generation balance, which varies from 2014 to 2022, as shown in Table 1. The monthly energy consumption for induction cookstoves is 96 kWh for a typical Ecuadorian family (Riofrio, 2015), so the monthly primary energy consumption for a family cooking with induction stoves was 175.97 kWh in 2014, and will be reduced to 120.74 kWh in the projected 2022 scenarios. The $CO_2$ emissions associated to the energy consumption for cooking for a typical Ecuadorian household using induction stove was 0.47 $tCO_2$ per year in 2014, and 0.14 $tCO_2/y$ in 2022, while for LPG (using 0.23$tCO_2$/MWh from IDAE, 2012) is 0.75 $tCO_2/y$.

**Table 1**
Primary energy for electric generation for 2014 and 2022.

| | Current situation (2014) | | | | | | Future scenario (2022) | | | | | |
|---|---|---|---|---|---|---|---|---|---|---|---|---|
| Source | Energy [MWh] | % | $C_i$ to Pe | Pe [MWh] | Emission factor [tCO2/MWh] | tCO2 | Energy [MWh] | % | $C_i$ to Pe | Pe [MWh] | Emission factor [tCO2/MWh] | tCO2 |
| Biomass Gen. | 399471.18 | 1.64 | 3.04 | 1214392.4 | 0 | 0 | | | | | 0 | 0 |
| Solar Gen. | 16482.7 | 0.07 | 1 | 16482.7 | 0 | 0 | | | | | 0 | 0 |
| Wind Gen. | 79742.47 | 0.33 | 1 | 79742.4 | 0 | 0 | | | | | 0 | 0 |
| **Total Renewable Gen.** | 495696.3 | 2,04 | [a]2.64[b] | 1310617.4 | 0 | 0 | 553000 | 1.30 | [a]2.64 | 1459920 | 0 | 0 |
| **Hydroelectric Gen.** | 11457895.6 | 47,14 | 1 | 11457895.6 | 0 | 0 | 35729000 | 83.67 | 1 | 35729000 | 0 | 0 |
| Fuel oil Gen. | 5483600.4 | 22.56 | 2.77 | 15189573.1 | 0.8 | 4386880.3 | | | | | | |
| Gas natural Gen | 2964552.7 | 12.20 | 1.95 | 5780877.8 | 0.8 | 2371642.2 | | | | | | |
| Diesel Gen. | 2759169 | 11.35 | 2.77 | 7642898.1 | 0.8 | 2207335.2 | | | | | | |
| Crude oil Gen. | 1146299.3 | 4.72 | 2.77 | 3175249.1 | 0.8 | 917039.4 | | | | | | |
| **Total Thermoelectric Gen.** | 12353621.4 | 50.82 | [a]2.57 | 31788598.1 | 0.8 | 9882897.1 | 6420000 | 15.03 | [a]2.57 | 16520078.8 | 0.8 | 5136000 |
| **Total** | 24307213.3 | 100 | [a]1.83 | 44557111.1 | [a]0.41 | 9882897.1 | 42702000 | 100 | [a]1.26 | 53708998.8 | [a]0.12 | 5136000 |

[a] Weighted arithmetic mean.
[b] Includes losses in the transformations and transportation.

**Table 2**
Summary of micro- and macro-economic scenario in terms of LPG price (US$/15 kg cylinder), electricity cost to the user (US$ / kWh), electricity cost to the government (US$ / kWh) and subsidy for the initial 80 kWh of electricity consumed (US$ / kWh).

| | Baseline (2014) situation | BAU 2022 | New Polices 2022 |
|---|---|---|---|
| LPG (15 kg) price for user (US$) | 1.60 | 1.60 | 20.00 |
| Subsidy for LPG (15 kg) (US$) | 18.40 | 18.40 | 0 |
| **Total cost LPG (15 kg) (US$)** | **20.00** | **20.00** | **20.00** |
| Electricity cost to the user (US$ / kWh) | 0.092 | 0.0858 | 0.0858 |
| Subsidy to electricity (US$/kWh) | 0.070 | −0.0243 | −0.0243 |
| **Total electricity cost (US$ / kWh)** | **0.162** | **0.0615** | **0.0615** |
| First 80 kWh cost for induction stoves users (US$ / kWh) | 0 | 0 | 0.036 |
| Over 80 kWh cost for induction stoves users (US$ / kWh) | 0.092 | 0.0858 | 0.0858 |
| Population (p) | 16026220,3 | 18044656 | 18044656 |
| Number of households | 3457375 | 4829599 | 4829599 |
| Number of LPG stoves | 3407375 | 4759570 | 140804 |
| Number of induction stoves | 50000 | 70029 | 4688795 |

### 2.3. Users and governmental economic scenarios

The economic impacts on LPG and induction-using consumers and on the finances of the Ecuadorian government were analyzed to evaluate the NECP in Ecuador. For this purpose, the subsidies for cooking energy in 2014 and potential situations for 2022 have been considered. Currently, electricity prices to consumers do not represent actual cost, because prices are influenced by the direct and indirect state subsidies for fossil fuels (representing 50.82% of electricity generation (see Table 1) that have been and continue to be implemented. To evaluate the potential benefits of massive migration to induction cooking in Ecuador, the 2014 baseline and two potential future scenarios have been considered. The first future scenario is a business as usual for 2022 (BAU-2022), in which subsidies are offered but there is no active effort to wean consumers off subsidized LPG. The second future scenario considers the full complement of new policies and regulations in the context of NECP for 2022, including the elimination of LPG subsidies.

For 2022, both potential scenarios assume the transition to a more hydroelectric-focused energy balance in accordance with the Electrification Master Plan 2013–2022 (CONELEC, 2013). The detailed considerations for each scenario follow, and the summary is presented in Table 2.

#### 2.3.1. Baseline situation

Consumer LPG is subsidized in Ecuador, with a 15 kg cylinder costing US$ 1.60 at official distribution points and slightly more elsewhere. Meanwhile, the market cost in the region for 15 kg cylinders of LPG without subsidies is about US$ 20.00 (Riofrio, 2015). The cost of subsidized electricity for the household connections in Ecuador currently is 0.092 US$/kWh, based on data from ARCONEL

(ARCONEL, 2014). Meanwhile, the current production cost of the electricity to the Ecuadorian government is 0.162 US$/kWh (CONELEC, 2015). Much of this production cost is due to subsidies for fossil fuels, totaling US$ 1.016 billion in 2014 (CONELEC, 2013). Currently, as part of the NECP, induction stove adopters receive a 100% subsidy on the initial 80 kWh consumed, and pay 0.092 US$ / kWh for additional consumption (MEER, 2013). According to the Ministry of Industries (MIPRO in Ecuador), 50000 induction stoves had been sold through December 2014 (MIPRO, 2015), and literature sources based on census data cite 3407375 households using LPG as their primary cooking fuel for the same period (Riofrio, 2015).

*2.3.2. BAU 2022 scenario*

The technology migration policy of the NECP is discarded, but government efforts to change the energy balance continue as-planned. The subsidy and consumer cost of LPG is maintained as in "Baseline situation". The operation of new hydroelectric power plants will reduce the real cost of electricity to 0.0615 US$/kWh (CONELEC, 2013), but users will pay 0.0858 US$/kWh (ARCONEL, 2014). The 100% subsidy for the initial 80 kW is kept as in "Baseline situation" for induction stoves adopters, while the cost for consumption of electricity beyond this amount is 0.0858 US$/kWh. Growth projections for both the number of LPG stove and induction stove-using households have been made according to the INEC's projections for increase in total population and decrease in number of family members per household through 2022 (INEC, 2010).

*2.3.3. New Polices 2022 scenario*

The NECP is successfully implemented with near-total migration to induction cooking. All consumer subsidies for LPG are eliminated, and consumer prices remain at approximately their 2014 market levels (US $20/15 kg cylinder). Data from ARCONEL (2014), estimates that the cost of electricity to the Ecuadorian user will be 0.0858 US$/kWh. Meanwhile, the cost of the electricity to the Ecuadorian government will be 0.0615 US$/kWh, without considering investment costs, as according to the predictions published by CONELEC (CONELEC, 2013). In addition, after the NECP's initial 100% subsidy period for induction users ends, users will pay 0.04 US$/kWh for the initial 80 kWh (MEER, 2013), and 0.0858 US$/kWh for additional consumption. This scenario estimates that 4688795 induction stoves will be in-use by 2022, using the same population growth projections as the above-described scenarios.

## 3. Results

*3.1. Costs to the government and $CO_2$ emissions*

These results for the three scenarios described above in terms of global energy consumption are summarized in Table 3. In 2014, Ecuador consumed approximately 11,500 GWh of primary energy for cooking activities. The government's annual subsidy spending reached US$ 882 million for bottled LPG, and US$ 8 million for induction stoves. In the 2022 BAU scenario, where policies to promote induction cooking are limited to new electricity subsidies and the subsidy for LPG remains as it is currently, the energy subsidy costs to the state would increase to around US$ 1.233 billion annually for LPG, and nearly US$ 4 million for induction stoves. The reduction in the subsidy cost from US$ 8 to 4 million between 2014 and 2022, despite an increase in the number of induction stoves, is due to the reduction in real electricity costs as a consequence of the inauguration of the projected hydropower capacity (CONELEC, 2013), and because consumer electricity costs will exceed the cost of production in 2022 (see Table 3). In the BAU 2022 scenario, the primary energy consumption for cooking would increase to 16,000 GWh annually. Under the new policy scenario for 2022, in which the NECP successfully leads to the adoption of 4688795 induction stoves, the primary energy consumption for cooking would drop to around 7200 GWh annually. In this case, the budget for energy subsidies for cooking fuels would decrease to US$ 75 million annually, delivered solely through subsidized electricity prices.

The $CO_2$ emissions associated to the energy consumption for cooking was 2.5 million of $tCO_2$ in 2014, 99% of them linked to the LGP stoves. For the 2022 BAU scenario the $CO_2$ emissions increase 39% respect than 2014, similar to the increment of the number of households (40% higher) due the low penetration of induction stoves and despite the lower $CO_2$ emission of the new electricity mix. The new policies scenario decrease the $CO_2$ emissions associated to the energy consumption for cooking to 0.75 million of $tCO_2$, representing the 30% of the 2014 emissions, despite the increment of households over the 8-year period.

*3.2. Costs to the households*

The monthly and annual energy consumption of a typical Ecuadorian family under the current and both future situations are presented in Table 4, disaggregated by cooking technology. The cost analysis for families in the baseline situation gives a monthly energy cost of US$ 22.53 for LPG and US$ 17.66 for induction stove electricity. Under the baseline framework the state subsidizes US$ 259.07 for LPG, or US$ 168.96 for induction stove electricity per family per year. Removing subsidies for LPG in 2022, and projecting the success of the NECP (MEER, 2013), a family would spend US$ 281.60 for LPG or US$ 54.87 for induction cooking yearly. For the BAU 2022 scenario, where the subsidies for LPG and induction electricity are maintained and the real electricity cost is reduced, LPG users would pay the same amount as in 2014 and induction cookstove users would pay US$ 16.47 per year.

## 4. Discussion

Ecuador is changing its energy balance, seeking to triple hydroelectric generation capacity, and halve the thermoelectric generation based on fossil fuels. Total electricity generation will increase 75% from 2014 to 2022 while the total primary energy associated with this generation will increase only 20.5%. The conversion factor to primary energy for the electricity mix is reduced from 1.83 to 1.26 due the higher efficiency of hydroelectric plants. This is reflected also in the $CO_2$ conversion factor for the electricity mix, that drops from 0.41 to 0.12 $tCO_2$/MWh from 2014 to 2022.

Primary energy consumption for cooking in Ecuador will be reduced to 54% of its 2014 baseline value by 2022 if the NECP is implemented successfully, while in a BAU scenario it will increase 40%. Thus, the successful implementation of NECP implies only 38% of the primary cooking energy consumption vs. the BAU scenario.

The reduction on $CO_2$ emissions associated to energy for cooking is around 1.8 million tCO2/y in 2022, if NECP is successfully implement, respect 2014. This $CO_2$ reduction offers an opportunity to the government to offer 1.8 million $tCO_2$ on international carbon market, which revenues could reinforce the NECP implementation and its sustainability.

The current total energy cost for cooking in Ecuador is US$ 968 million per year (92% of it subsidized by government), which will increase to US$ 1.345 billion by 2022 if NECP is not implemented (92% subsidized), but will be reduced to US$ 372 million with full implementation (20% subsidy). The new policies scenario would save US$ 1.162 billion in annual government expenditure on cooking subsidies compared to the BAU situation. A single years of these savings is greater than the planned investment in induction stoves through the NECP, and equals 10% of the government's planned total investment in new hydroelectric power stations, new grid transmission infrastructure and induction stoves.

Based on 2014 data, the Ecuadorian government subsidizes US$ 55.60 per capita in cooking energy yearly, mainly in the form of

**Table 3**
Global cooking energy consumption, cost subsidies and $CO_2$ emissions for 2014 and 2022 scenarios.

|  | Units | 2014 | | 2022 | | | |
|---|---|---|---|---|---|---|---|
| Population | [U] | 16026220,3 | | 18044656 | | | |
|  |  | Current situation | | BAU | | New policies | |
|  |  | LPG | Induction | LPG | Induction | LPG | Induction |
| N° of Stoves | [U] | 3407375 | 50000 | 4759570 | 70029 | 140804 | 4688795 |
| Monthly consumption | [GWh/y] | 904.28 | 4.80 | 1263.14 | 6,72 | 37,37 | 450,12 |
| Annual consumption | [GWh/y] | 10851.40 | 57.60 | 15157.71 | 80,67 | 448,41 | 5401,49 |
| Primary energy | [GWh/y] | 11393.97 | 105.59 | 15915.59 | 101,46 | 470,83 | 6793,79 |
| Total Cost | [MUS$/y] | 959.52 | 9.33 | 1340.29 | 4,96 | 39,65 | 332,19 |
| Subsidy cost | [MUS$/y] | 882.76 | 8.45 | 1233.07 | 3,81 | 0 | 74,90 |
| $CO_2$ Emissions | [tCO2/y] | 2540086.5 | 23419.2 | 3548103.6 | 9703.0 | 104964.8 | 649666.6 |
| Total primary energy | [GWh/y] | 11499.56 | | 16017.30 | | 7264.62 | |
| Total cost | [MUS$/y] | 968.85 | | 1345.26 | | 371.84 | |
| Total subsidy cost | [MUS$/y] | 891.20 | | 1236.88 | | 74.90 | |
| Total $CO_2$ Emissions | [tCO2/y] | 2563505.7 | | 3557806.6 | | 754631.3 | |

**Table 4**
Monthly and annual energy consumption scenario for a typical ecuadorian family for both technologies.

|  | Units | 2014 | | 2022 | | | |
|---|---|---|---|---|---|---|---|
|  |  | Current situation | | Business as Usual | | New policies | |
|  |  | GLP | Induction | GLP | Induction | GLP | Induction |
| Monthly comsumption | [kWh] | 265.39 | 96 | 265.39 | 96 | 265.39 | 96 |
| Annual comsumption | [kWh] | 3184.68 | 1152 | 3184.68 | 1152 | 3184.68 | 1152 |
| Total cost per year | [US$] | 281.50 | 186.62 | 281.60 | 70.85 | 281.60 | 70.85 |
| Subsidy cost per year | [US$] | 259.07 | 168.96 | 259.07 | 54.37 | 0 | 15.97 |
| User cost per year | [US$] | 22.53 | 17.66[a] | 22.53 | 16.47[b] | 281.60 | 54.87[c] |

[a] The first 80 kWh are not paid and above the rate is 0.092 US$.
[b] The first 80 kWh are not paid and above the rate is 0.0858 US$.
[c] The first 80 kWh is $0.04 and above the rate: 0.0858 US$.

imported LPG. If new policies are not undertaken, the BAU 2022 scenario demonstrates that this amount will increase to US$ 68.50, with continued dependence on imported fuel. However with the implementation of the NECP, the New Policies 2022 scenario would reduce the subsidy to US$ 4.10 per year per capita.

Although the government will reduce the expenditure on cooking energy subsidies 91% from the baseline through the successful implementation of the NECP by 2022, consumers would see their costs increase. Continuing consumers of LPG will see costs increase from US$22.53 to 281.70 by 2022 (12.5 times higher), while induction users will have their costs increased from US$17.66 to 54.87 (3.1 times higher). However, a majority of families currently cook with LPG, so the price increase with the massive transition to induction will imply cooking energy costs 2.4 times higher than baseline expenditures.

To provide context to the micro-economic analysis for Ecuadorian families, it is essential to understand that the annual minimum wage in the country for the baseline period was US$ 4080.00, while the cost of the government-defined basic goods package was US$ 7548.12 (INEC, 2014). For the baseline period, cooking with LPG stoves represented 0.29% of the cost of the basic goods package, and 0.55% of the minimum wage, while cooking with induction stoves represented 0.23% and 0.43% respectively. The minimum wage and basic good costs are expected to rise by 2022, but for reference calculations they have been considered fixed. With this consideration, the increase of cooking energy costs for a family in 2022 using induction stoves, with the NECP in-place will imply 0.73% of the cost of the basic goods package and 1.34% of the minimum wage, while cooking with LPG stoves will imply 3.73% of basic goods costs and 6.90% of the minimum wage.

## 5. Conclusions and policy implications

This study's analysis of technological migration for cooking in Ecuador, considering the NECP and the projected inauguration of new hydropower capacity in the country, demonstrates the major savings available to the Ecuadorian state through the implementation of new policies. If the LPG subsidy is maintained, by 2022 state spending on cooking energy subsidies will increase 39% vs. its 2014 baseline level, reaching over US$ 1.2 billion annually. Over the same period, the massive introduction of induction stoves and migration from LPG foreseen in the NECP's policies would lead to a 92% reduction in subsidy spending. Yearly savings generated by the program are equal to 10% of the government's total investment in current efforts to green and modernize the energy balance, improve transmission infrastructure, and subsidize and promote induction cookstoves. Also, the 1.8 million of tCO2/y reduction, associated to energy for cooking, could be sell in international carbon market, improving the macroeconomics of the NECP.

While government spending could be significantly reduced, the average Ecuadorian family would experience an increase in its expense on cooking energy under the NECP. With the planned elimination of the subsidy for LPG before 2022, families who do not migrate to induction cooking would see their cooking energy costs multiplied more than twelve-fold as compared to 2014 levels. However, migration to induction offers an affordable alternative for families, with a cost increase of only 2.4 times between 2014 LPG expenses and 2022 electricity costs.

Ecuador's implementation of next-generation clean cooking technology through the introduction of induction stoves powered largely by new hydropower facilities looks to replace the often-abused subsidy for LPG. The substitution of this largely imported fossil fuel would lead to reduced government expenditure, greater energy sovereignty, improvements in health and security for families (mainly women and children), and represents a feasible option for ending the current LPG subsidy. This program shares the economic impact of major technological migration with the country's population, without generating the financial hardships that an abrupt end to the LPG subsidy could cause.

## Acknowledgments

The authors acknowledge the Secretaría Nacional de Planificación y Desarrollo (SENPLADES) for financing the implementation of the present research. This work was sponsored by the Prometeo project of the Secretaria de Educación Superior, Ciencia, Tecnología e Innovación (SENESCYT), hosted by the Republic of Ecuador. The information necessary to complete this work was provided by the Ministerio de Electricidad y Energía Renovable (MEER) of Ecuador. The authors wish to acknowledge the revision of the document by Samay Schütt and Samuel Schlesinger, and the suggestions from the anonymous reviewers. Jaime Martí-Herrero acknowledge the financial support to CIMNE via the CERCA Programme / Generalitat de Catalunya.